\documentclass[prl,a4paper,showpacs,reprint]{revtex4-1}

\pdfoutput=1

\usepackage{amsmath}
\newcommand{\nvec}{\boldsymbol{n}}
\newcommand{\mvec}{\boldsymbol{m}}
\newcommand{\kvec}{\boldsymbol{k}}
\newcommand{\qvec}{\boldsymbol{q}}
\newcommand{\rvec}{\boldsymbol{r}}

\newcommand{\deltavec}{\boldsymbol{\delta}}

\usepackage{graphicx}

\begin{document}


\title{Polarons in highly doped atomically thin graphitic materials.}

\author{J.P. Hague}
\affiliation{The Open University, Walton Hall, Milton Keynes, MK7 6AA, United Kingdom}

\begin{abstract}
Polaron spectral functions are computed for highly doped
graphene-on-substrate and other atomically thin graphitic systems
using the diagrammatic Monte Carlo technique. The specific aim is to
investigate the effects of interaction on spectral functions when the
symmetry between sub-lattices of a honeycomb lattice has been broken
by the substrate or ionicity, inducing a band gap. Introduction of
electron-phonon coupling leads to several polaronic features, such as
band-flattening and changes in particle lifetimes. At the K point,
differences between energies on each sub-lattice increase with
electron-phonon coupling, indicating an augmented transport gap, while
the spectral gap decreases slightly. Effects of phonon dispersion and
long-range interactions are investigated, and found to lead to
only quantitative changes in spectra.
\end{abstract}

\date{5th July 2012}

\maketitle

\section{Introduction}

The graphene hexagonal lattice leads to exceptional electronic
properties: a zero bandgap semiconductor with very high mobilities,
and essentially massless Dirac fermions \cite{castroneto2009a}. This
has already led to a number of applications, but transistors for
digital applications remain elusive because they require a gap. While
a gap cannot be induced in suspended monolayer graphene, recent
experimental work using angle resolved photo-emission spectroscopy
(ARPES) has shown that there may be bandgaps in graphene on certain
substrates: ARPES measurements have found a gap in graphene on a
monolayer of intercalated gold on ruthenium \cite{enderlein2010a} and
there has been significant debate regarding whether a gap is present
in monolayer graphene on silicon carbide
\cite{zhou2007a,bostwick2007a}. 

The debate about the existence of the gap seen for graphene on SiC
relates to the interpretation of ARPES measurements \footnote{It is
  also worth noting that the physics of the graphene on SiC system may
  be complicated because of the nature of the surface reconstruction
  \cite{qi2010a}.}. The authors of Ref. \onlinecite{zhou2007a} claim
that there is a gap of around 0.26eV because of modulation of the
potential fue to the substrate, however it has been claimed that this
gap is a mis-interpretation of other excitations such as polarons and
polaritons \cite{bostwick2007a,rotenberg2008a}.  ARPES measurements
are carried out using either doping or gating to move the system away
from half filling. For example, in Ref. \onlinecite{zhou2007a} the
Fermi energy, $E_F$ is shifted by around 0.4eV. Bostwick {\it et al.}
\cite{bostwick2007a} consider a system where $E_F$ is shifted by
0.45eV. Where evidence has been found for the opening of the gap in
the ruthenium system due to a breaking of the symmetry of the two
carbon sub-lattices in graphene, the Fermi energy was reported to be
around 0.15eV below the Dirac point \cite{enderlein2010a}. Therefore,
the effects of doping on spectral functions are of interest.

An alternative material with a honeycomb lattice and a bandgap
resulting from a modulated potential is boron nitride (BN), which can
be mechanically exfoliated in atomically thin layers
\cite{novoselov2005a}. Atomically thick hexagonal BN (h-BN) bonds
through $sp_2$ hybridization (just as in graphene) and has itinerant
electrons in $\pi$ orbitals \cite{alem2009a}. In BN, ionicity means
that $\pi$ orbitals on N sites are shifted up in energy by $+\Delta$,
with a decrease in energy of $-\Delta$ on B sites, causing a gap of
order $2\Delta$. {\it Ab-initio} simulations have established hopping
in the BN monolayer to be $t=2.33$eV \cite{ribeiro2011a} and that the
parameter, $\Delta=1.96$eV$=0.84t$. Experiments on monolayer h-BN find
a gap of $5.56$eV \cite{song2010a}. Longitudinal acoustic (LA) phonons
energies peak at around 140meV at the M point, and transverse acoustic
(TA) phonons at around 110meV at the K point. Optical phonon energies
range between 160meV and 200meV \cite{serrano2007a}, and strong
coupling between electrons and phonons is expected because individual
sites have an net charge in the ionic materials.

Several other atomically thin materials can be mechanically
exfoliated, including SnS$_2$, CdI$_2$ and MoS$_2$, but these have the
chalcogenide structure, rather than the graphene honeycomb
lattice. Other layered materials that have a honeycomb structure
include GaN, which has a bandgap of 2.15eV and can be grown in thin
films \cite{petalas1995a}. The related AlN can also exist in a
hexagonal structure with bandgap 6.28eV
\cite{litimein2002a}. Following the discovery of silicene
\cite{padova2010a}, it is possible that other III-V semiconductors can
be encouraged to grow in thin hexagonal films.

Recently, I calculated that gaps caused by modulated potentials on
honeycomb lattices may be enhanced by introducing strong
electron-phonon coupling through a highly polarizable superstrate
\cite{hague2011b,hague2012a}.  Effective electron-electron
interactions can be induced via a strong interaction between the
electrons in a graphene monolayer and phonons in a strongly
polarizable substrate because of limited out of plane screening,
similar to that seen for quasi-2D materials
\cite{alexandrov2002a}. Such interactions have been experimentally
demonstrated between carbon nanotubes and a SiO$_2$ substrate
\cite{steiner2009a}, and are necessary to account for the lower
mobilities of graphene on SiO$_2$ \cite{fratini2008a}. These
interactions will form polaronic states and affect the overall
electronic structure in the graphene monolayer. For the
graphene-on-substrate systems with small gaps, devices could be
covered with highly polarisabe superstrates to combine gap effects with
strong electron-phonon interaction.

No work has previously been carried out to establish the properties of
polarons in a gapped graphene system, and this paper discusses how
spectral functions for a single electron could be affected by
electron-phonon interactions, as an approximation to polaronic effects
in the highly gated or doped regime. I calculate the effects of
electron-phonon interaction on electrons in a honeycomb lattice
(graphene) where a gap has been opened with a modulated potential. I
present results computed using the numerically exact diagrammatic
quantum Monte Carlo (DQMC) technique, and compute spectral functions
using stochastic analytic inference. The DQMC method used here is
valid for a single polaron - that is a single electron at the bottom
of an empty band interacting with a cloud of phonons. So, results
presented in this paper are approximately valid in the heavily doped
regime, well away from half-filling. The paper is organized as
follows: The model Hamiltonian is introduced in
Sec. \ref{sec:model}. Details of the extensions to DQMC specific to
graphene are explained in
Sec. \ref{sec:method}. Sec. \ref{sec:results} presents detailed
spectral functions for a range of model parameters. A summary and
conclusions are given is Sec. \ref{sec:conclusions}.

\section{Model Hamiltonian}
\label{sec:model}

\begin{figure}
\includegraphics[width = 70mm]{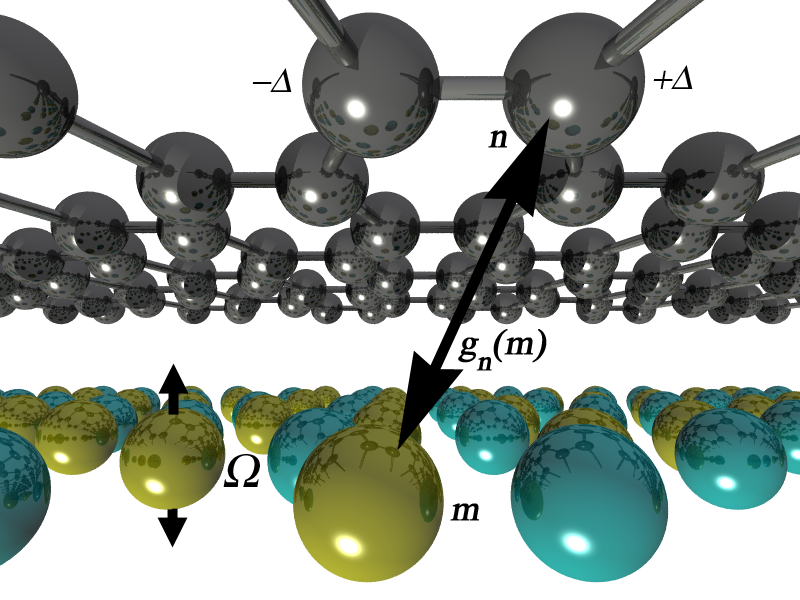}
\caption{(Color online) Graphene-substrate system annotated with interactions and sublattices. Electron-phonon interactions between the graphene layer and substrate are poorly screened, and large interactions are possible.}
\label{fig:schematic}
\end{figure}

\begin{figure*}
\includegraphics[width = \textwidth]{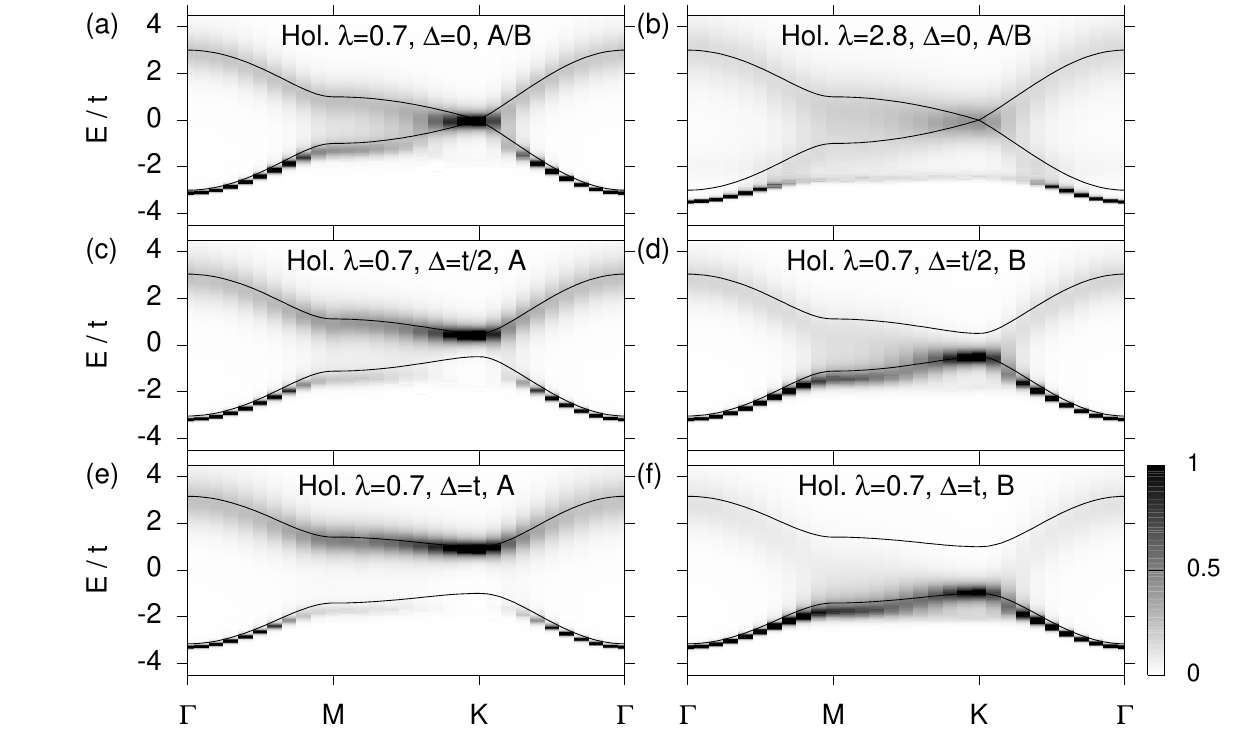}

\caption{Image plots of the graphene spectral function across the
  Brillouin zone. Various $\Delta$ and $\lambda$ are shown. The
  non-interacting dispersion is overlaid. At $\Delta=0$, A and B
  sub-lattices are symmetrical, leading to identical spectral
  functions, so only a single panel is shown for each
  $\lambda$. Comparison with the non-interacting dispersion shows that
  there is some flattening of the band close to the K point. The
  spectral function is sharp close to the K point. In the vicinity of
  the $\Gamma$ point, the spectral function is also sharp for states
  within an energy $\hbar\Omega$ of the bottom of the
  band. Quasi-particle lifetime (related to the inverse of the width)
  is greatly reduced close to the tops of the bands. Increase in
  $\Delta$ breaks the AB symmetry. Some band flattening is seen. There
  is also a weak excitation associated with B sites at higher
  energies, which touches the lower band. Panel (a) $\Delta=0$,
  $\lambda=0.7$, (b) $\Delta=0$, $\lambda=2.8$ where A and B site
  electrons are identical. (c) and (d) $\Delta=0.5t$, $\lambda=0.7$,
  and (e) and (f) $\Delta=t$, $\lambda=0.7$, showing A and B site
  electrons respectively. There is only a single electron, so the
  chemical potential is at the bottom of the band, $E_0$. The origin
  is arbitrary, but for consistency has been taken as the point where
  the unperturbed bands cross.}
\label{fig:vardeltaimage}
\end{figure*}

\begin{figure*}
\includegraphics[width = \textwidth]{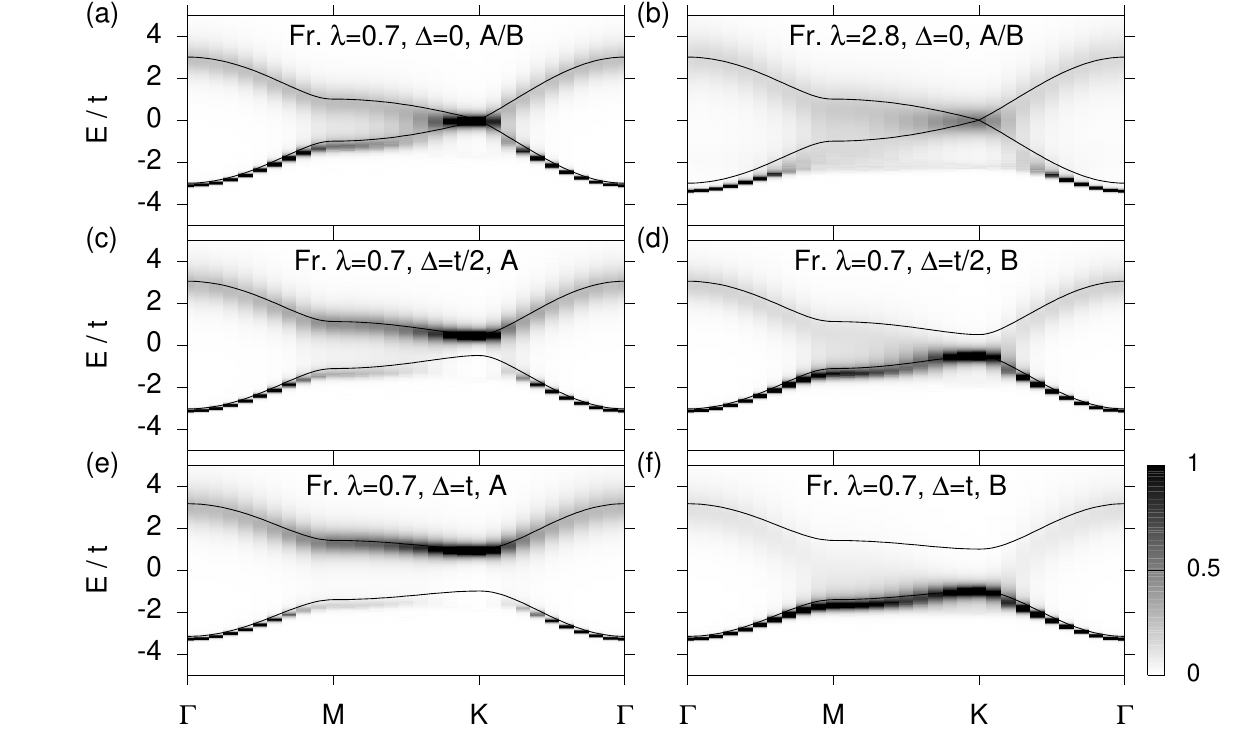}
\caption{As Fig. \ref{fig:vardeltaimage} with the leading nearest neighbor corrections from Fr\"ohlich interactions included. There are only minor
  changes to the results.}
\label{fig:vardeltaimagefroh}
\end{figure*}

\begin{figure}
\includegraphics[width = 75mm]{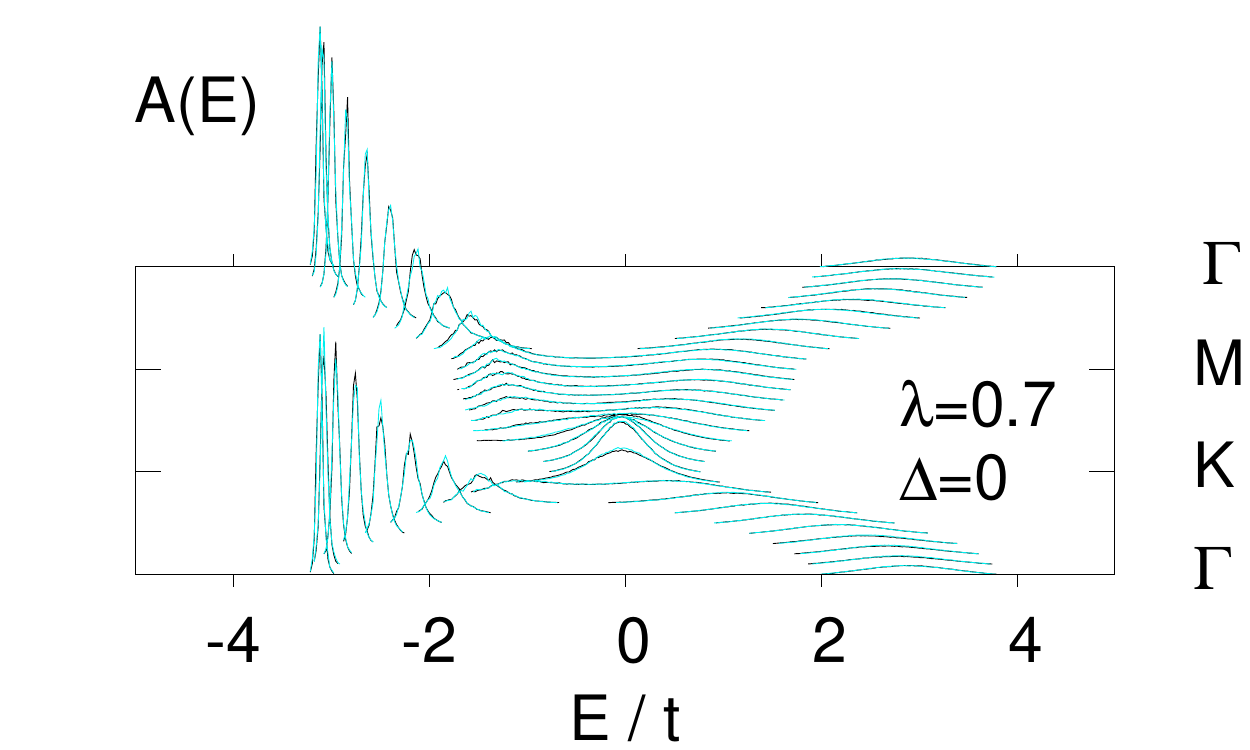}
\includegraphics[width = 75mm]{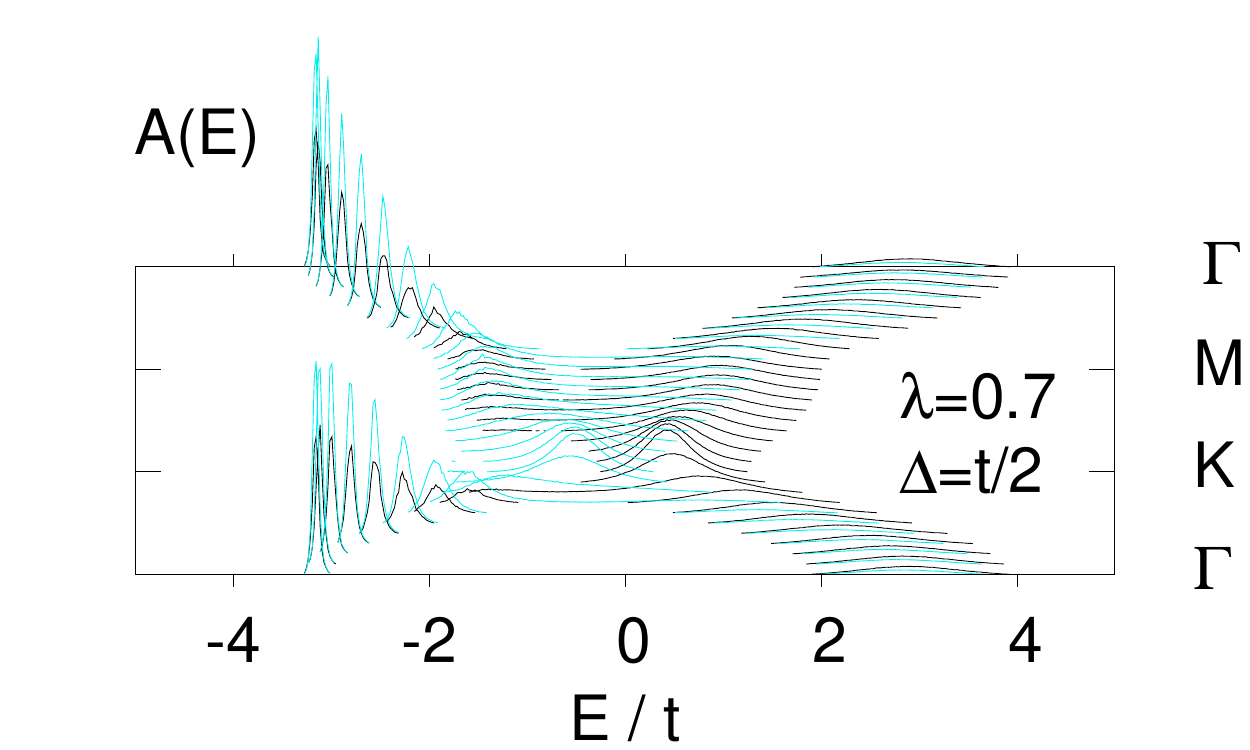}
\includegraphics[width = 75mm]{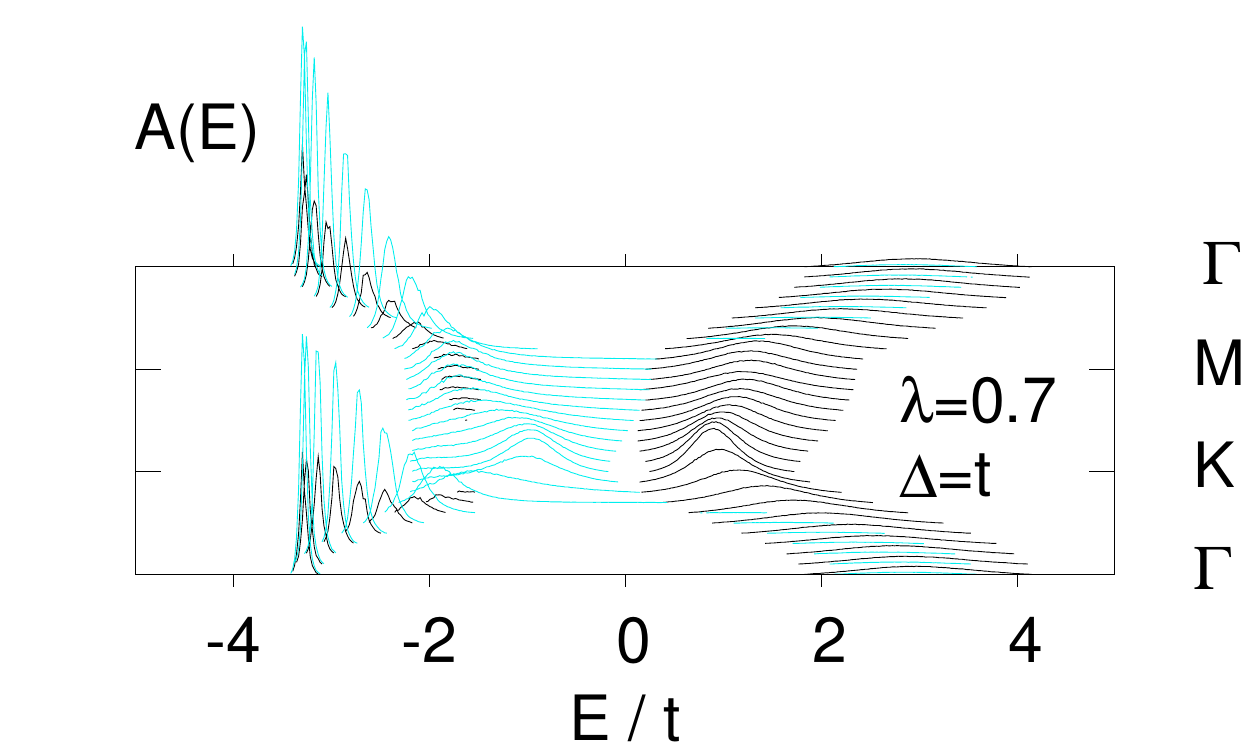}

\caption{(Color online) Variation of the graphene spectral function
  across the Brillouin zone for various $\Delta$ and $\lambda =
  0.7$. The same data as Fig. \ref{fig:vardeltaimage} is shown with
  both A and B sublattices superimposed on the same plot. From top to
  bottom, $\Delta=0$, $\Delta=t/2$ and $\Delta=t$. The gap opening is
  clearly visible. Broadening of the spectral functions can be seen,
  especially at larger energies.}
\label{fig:vardelta}
\end{figure}

\begin{figure*}
\includegraphics[width = \textwidth]{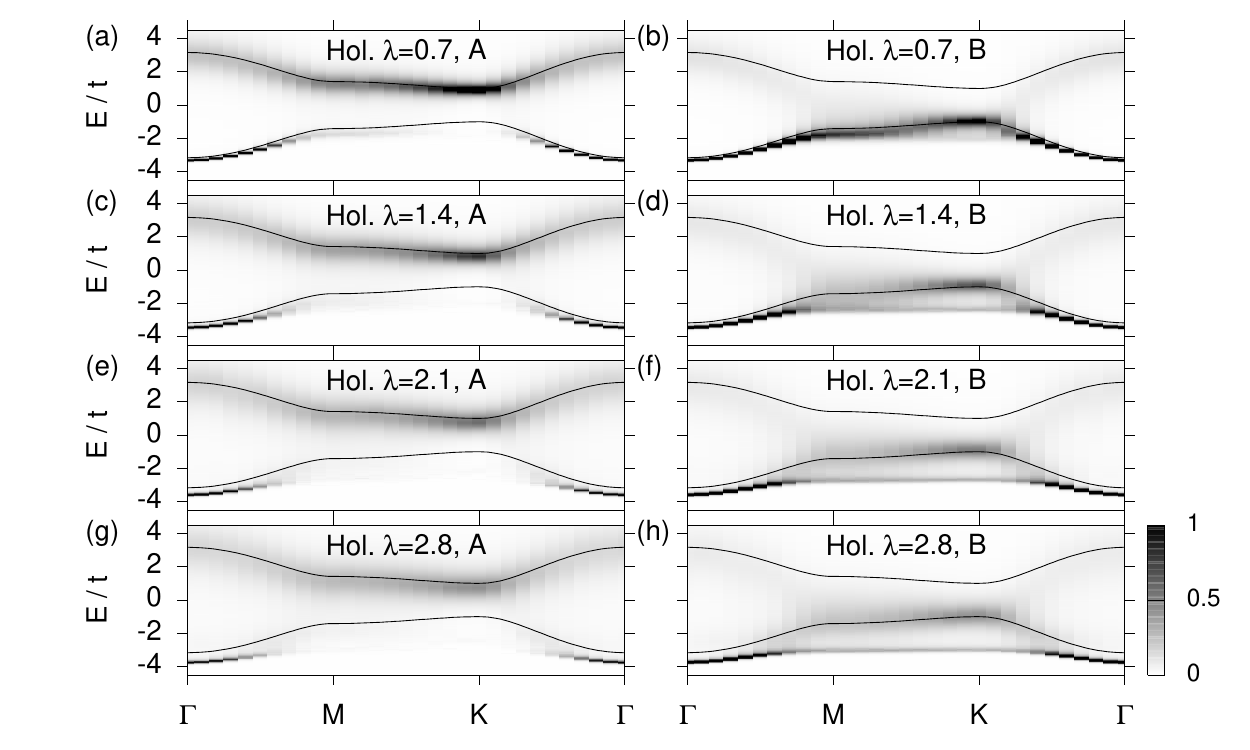}
\caption{Image plots of the graphene spectral function
  across Brillouin zone. $\Delta = t$ and various $\lambda$. Panels
  (a) and (b) $\lambda=0.7$ (c) and (d) $\lambda=1.4$, (e) and (f)
  $\lambda=2.1$ and (g) and (h) $\lambda=2.8$. A sites can be seen on
  the left and B on the right. A flat polaron band can be seen forming
  at an energy around $\hbar\Omega$ above the bottom of the band for
  very large $\lambda$. The quasi-particle lifetime decreases
  dramatically with increased $\lambda$.}
\label{fig:varlambdaimage}
\end{figure*}

\begin{figure}
\includegraphics[width = 75mm]{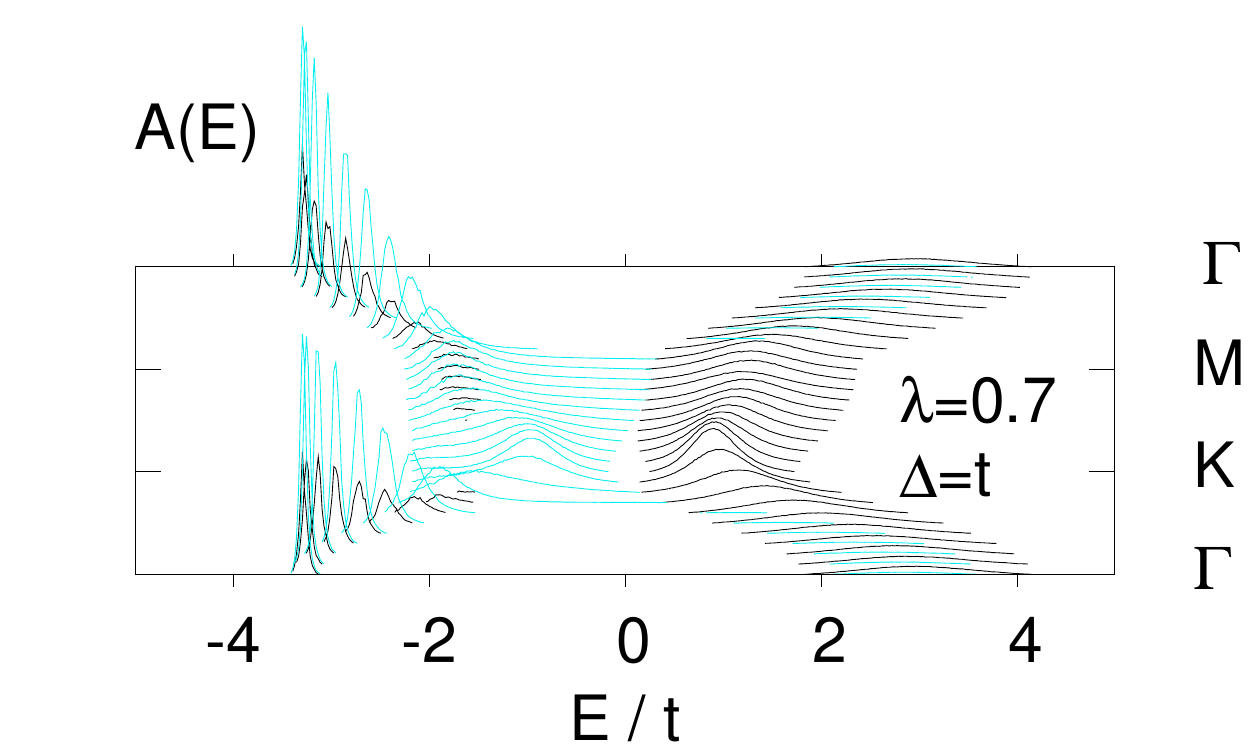}
\includegraphics[width = 75mm]{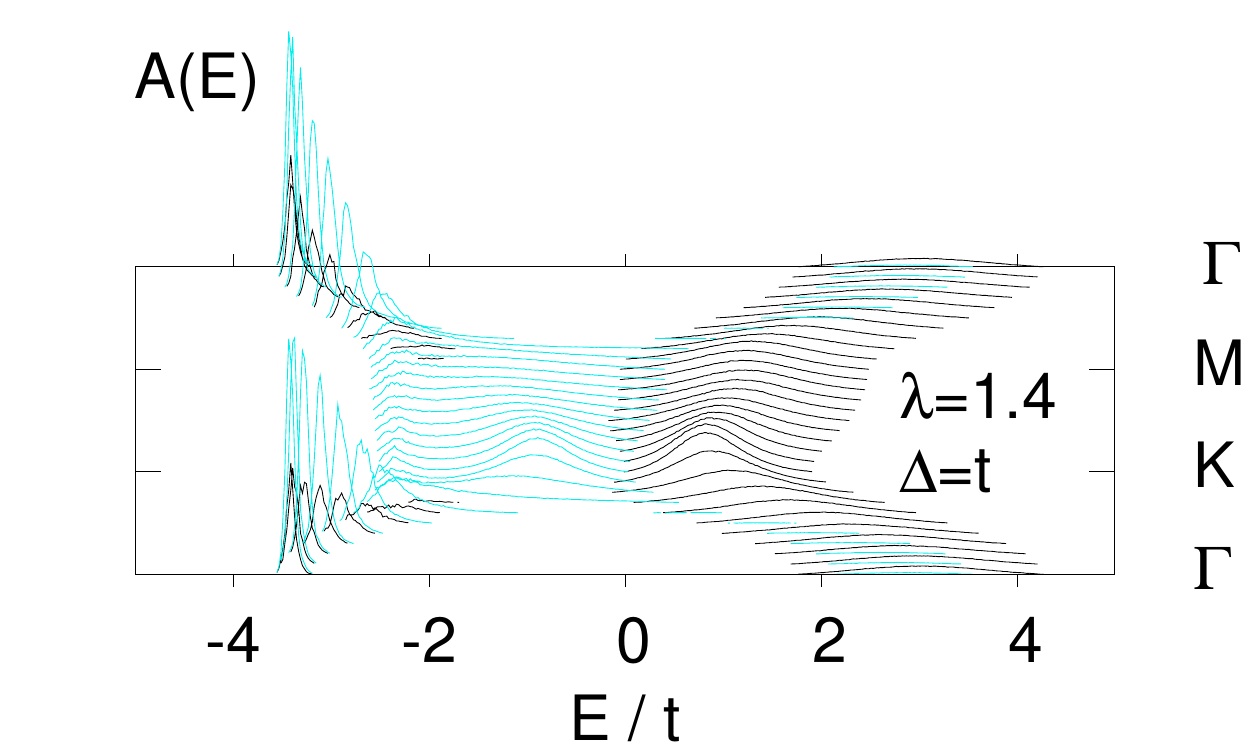}
\includegraphics[width = 75mm]{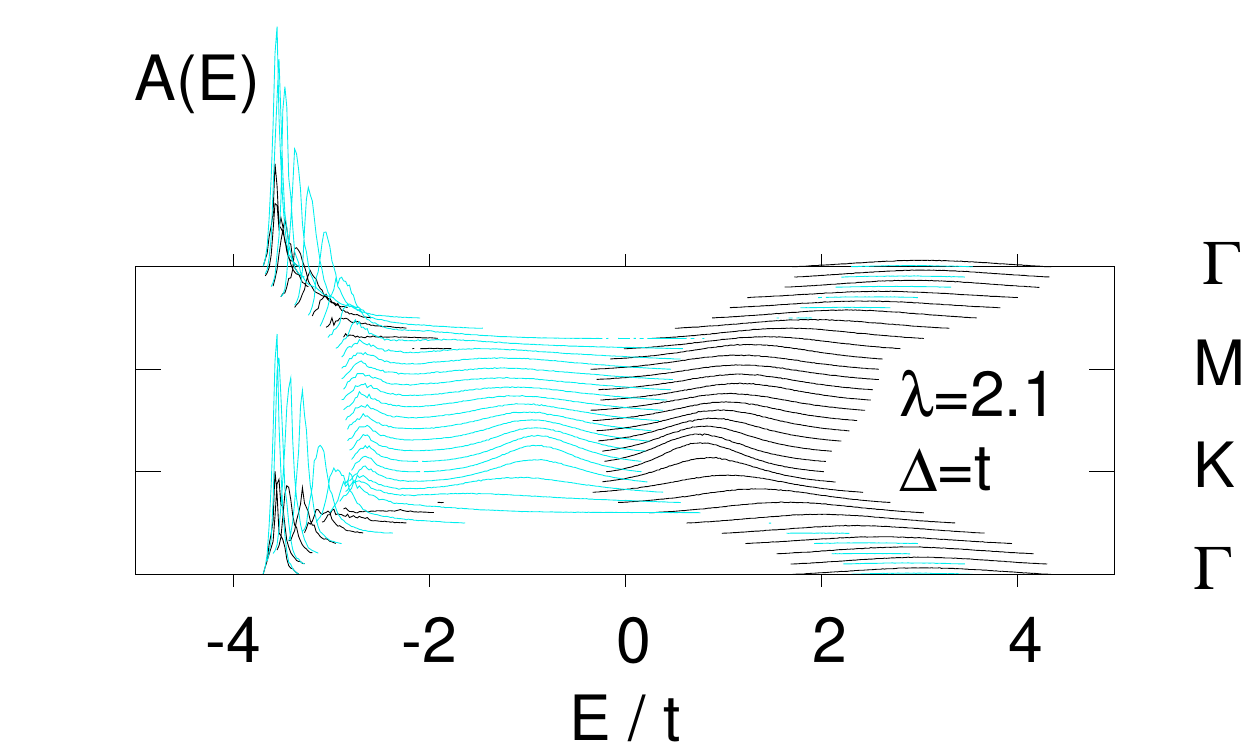}
\includegraphics[width = 75mm]{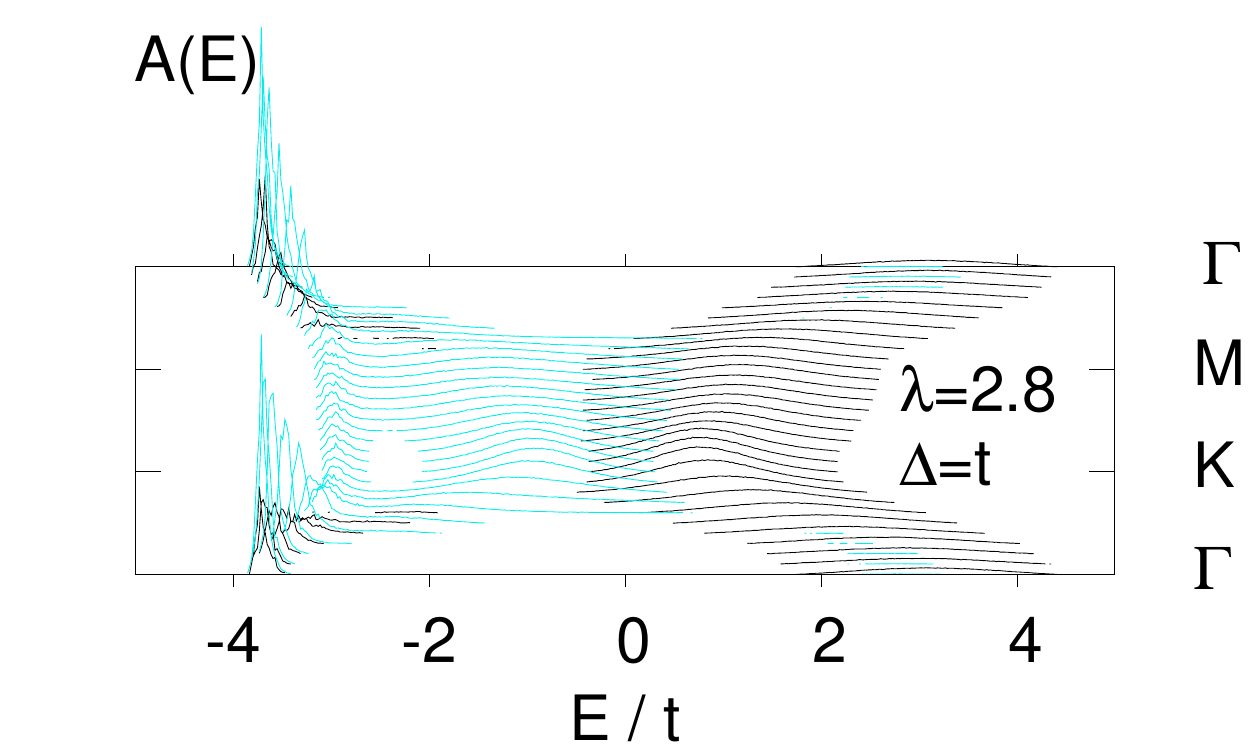}
\caption{(Color online) Graphene spectral function across Brillouin zone. $\Delta = t$ and various $\lambda$. Data is as Fig. \ref{fig:varlambdaimage} but with results for both sub-lattices plotted together. It can be seen that the gap is robust against increase in $\lambda$, although a number of additional excitations appear.}
\label{fig:varlambda}
\end{figure}

\begin{figure*}
\includegraphics[width = \textwidth]{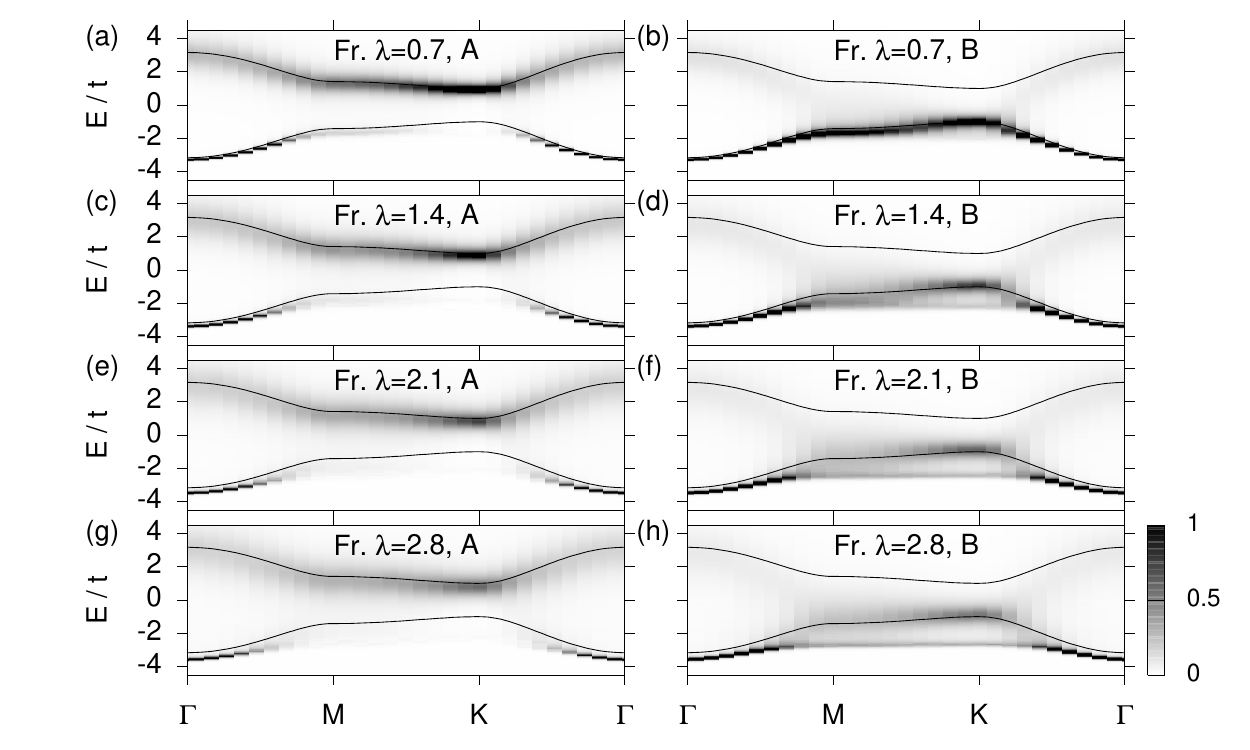}
\caption{As Fig. \ref{fig:varlambdaimage} with the leading corrections from
  Fr\"ohlich interactions included. There are only moderate
  changes to the results, consistent with a small reduction in effective $\lambda$.}
\label{fig:varlambdaimagefroh}
\end{figure*}

\begin{figure}
\includegraphics[width = 85mm]{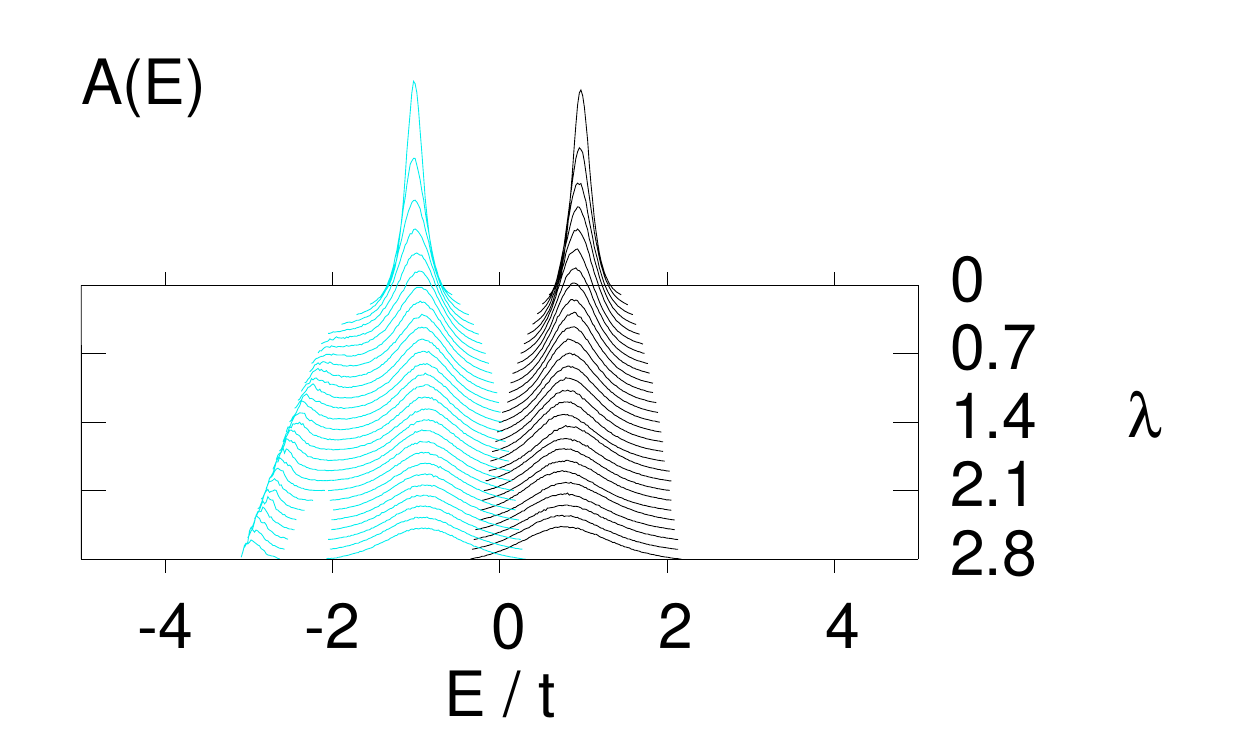}
\caption{(Color online) Graphene spectral function at the K
  point. $\Delta = t$ and $\lambda$ is varied. The spectral gap (seen
  at approximately $E=5t$) is robust. The spectral functions broaden
  with increased $\lambda$, and the polaron band rapidly drops in
  energy leading to an enhanced transport gap.}
\label{fig:diracvarlambda}
\end{figure}

\begin{figure*}
\includegraphics[width = \textwidth]{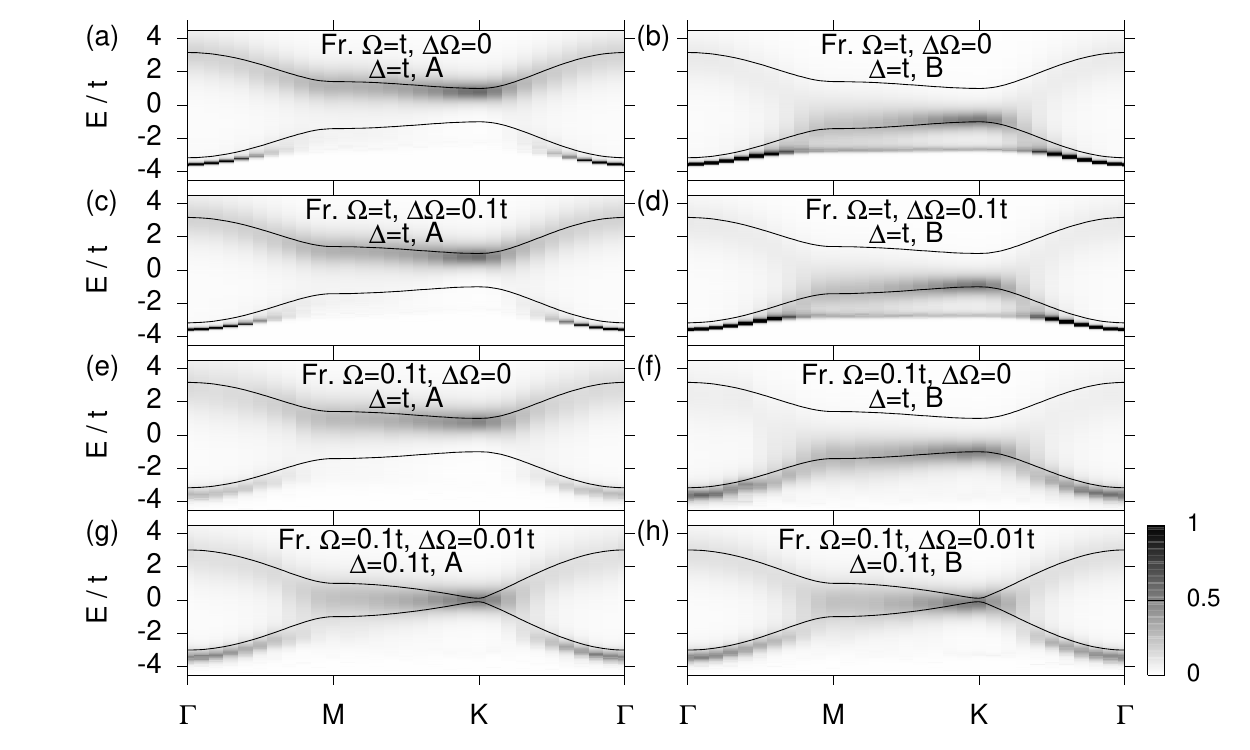}
\caption{Effects of dispersion, $\Delta\Omega$, and phonon frequency,
  $\Omega$ on the spectral function. $\lambda=2.8$
  throughout. Broadening the phonon dispersion has no major effects on
  the spectral function. On the other hand, decreasing the phonon
  frequency leads to a broading of all features corresponing to a
  sharp decrease in quasi-particle lifetime, and indicating a
  significant increase in scattering.}
\label{fig:dispersivephonons}
\end{figure*}

In this section, I introduce details of a model Hamiltonian for
polaronic effects in gapped systems. Three components are required to
examine polaronic interactions between graphene and surface phonons in
a substrate (or superstrate): (a) Intersite hopping within the
hexagonal plane, which is well known to properly account for the band
structure of monolayer graphene, (b) An electron-phonon interaction to
account for any polaronic effects from interaction with surface
phonons in the substrate, (c) Direct Coulomb interaction between the
electrons and substrate. The same components are valid for monolayers
of ionic graphitic materials such as BN.

A complication of the Hamiltonian required to describe graphene, is
that a basis of two atoms is needed to represent the honeycomb
lattice. This leads to a Hamiltonian with the form,
\begin{equation}
H_{\rm tb}  =  \sum_{\kvec}(\phi_{\kvec} a^{\dagger}_{\kvec}c_{\kvec} + \phi^{*}_{\kvec} c^{\dagger}_{\kvec}a_{\kvec})
\end{equation}
\begin{widetext}
\begin{equation}
H_{\rm e-ph} = \sum_{\kvec\qvec} g_{\qvec} \left[c^{\dagger}_{\kvec-\qvec}c_{\kvec}(d^{\dagger}_{\qvec} + d_{-\qvec}) + a^{\dagger}_{\kvec-\qvec}a_{\kvec}(b^{\dagger}_{\qvec} + b_{-\qvec})\right]
+\tilde{g}_{\qvec} \left[a^{\dagger}_{\kvec-\qvec}a_{\kvec}(d^{\dagger}_{\qvec} + d_{-\qvec}) + c^{\dagger}_{\kvec-\qvec}c_{\kvec}(b^{\dagger}_{\qvec} + b_{-\qvec})\right]
\end{equation}
\end{widetext}
\begin{equation}
H_{\rm ph}  =  \sum_{\qvec} \Omega_{\qvec} (b^{\dagger}_{\qvec} b_{\qvec} + d^{\dagger}_{\qvec} d_{\qvec})
\end{equation}
\begin{equation}
H_{\rm static}  =  \sum_{i\sigma}V(\rvec_{i})n_{i\sigma}
\end{equation}
Here, $H_{\rm tb}$ is the tight binding Hamiltonian representing the
kinetic energy of the electrons in the graphene monolayer,
$\phi_{\kvec}=-t\sum_i \exp(i\kvec.\deltavec_i)$, where $\kvec$ is the electron momentum and $\deltavec_i$ are the
nearest neighbor vectors from A to B sub-lattices,
$\deltavec_1=a(1,\sqrt{3})/2$, $\deltavec_2=a(1,-\sqrt{3})/2$ and
$\deltavec_3=(-a,0)$. Electrons are created on A sites with the
operator $a^{\dagger}_i$ and B sites with $c^{\dagger}_i$.

The term $H_{\rm e-ph}$ describes the electron-phonon
interaction. Phonons with momentum $\qvec$ are created on A sites with
$b^{\dagger}_{\qvec}$ and on the B sublattice with
$d^{\dagger}_{\qvec}$. Thus, interactions between electrons and
phonons on the same sub-lattices have magnitude $g_{\qvec}$, and
interactions between different sub-lattices have magnitude
$\tilde{g}_{\qvec}$. The Fr\"ohlich form for the electron-phonon
interaction has been demonstrated experimentally for carbon nanotubes
on SiO$_2$ \cite{steiner2009a}, is theoretically proposed for quasi-2D
systems where out of plane hopping is low \cite{alexandrov2002a} and
has also been found necessary to account for mobilities for graphene
on substrate systems \cite{fratini2008a}. The interactions between
electrons in the graphene and polarizable ions in the substrate are
shown schematically in Fig. \ref{fig:schematic}. There are only weak
interactions between electrons and phonons within the graphene plane,
accounting for the very high mobility of suspended graphene. In ionic
materials such as BN, in plane interactions may be stronger. Two
interactions are considered: A Holstein interaction, where the Fourier
transform of the force function, is momentum independent and
$\tilde{g}=0$, and a Fr\"ohlich interaction, where the Fourier
transform,
$g_{\qvec}=\sum_{\nvec}g_{0}[\nvec]e^{i\qvec\cdot\nvec}/\sqrt{N}$ of
the force function is truncated at nearest neighbors so
$\tilde{g}_{\qvec}\propto \exp(i\qvec.\deltavec_i)$.

$H_{\rm ph}$ is the energy of the phonons in the substrate (treated as
harmonic oscillators, and including both kinetic and potential energy
of the ions), and the phonon dispersion is
$\Omega_{\qvec}=\Omega+\Delta\Omega\left(\cos(q_x\sqrt{3})+2\cos(q_y\sqrt{3}/2)\cos(3q_x/2)\right)$,
where $\Delta\Omega$ controls the width of the phonon band. It
is usual to define a dimensionless electron-phonon coupling,
$\lambda=\Phi(0,0)/tM\Omega^2$, where the effective interaction
$\Phi(\nvec,\nvec') =
\sum_{\mvec}g_{\mvec}[\nvec]g_{\mvec}[\nvec']/(2M\Omega)$, where
$g_{\mvec}[\nvec]=\kappa/(1+|\nvec-\mvec|^2)^{3/2}$ and $\kappa$ is a
coupling constant. Typical effects of the electron-phonon interaction
in conventional semiconductors are the generation of polarons, leading
to changes in the bandstructure and thus the effective mass of the
carriers, modification of the Landau levels and changes in the optical
properties of the material such as absorption peaks in the mid
infra-red.

To complete the model of graphene on a substrate, $H_{\rm static}$
describes interaction between electrons in the monolayer and a static
potential, $V(\rvec_i)$, induced by the substrate (where $\rvec_i$ are vectors to lattice sites). Here, a modulated
potential is considered where A sites have energy $\Delta$ and B sites
$-\Delta$ leading to breaking of the symmetry between A and B
sub-lattices and giving rise to a gap. Such a form has been suggested
to explain gaps in graphene on SiC \cite{zhou2007a} and graphene on
rubidium \cite{enderlein2010a}, and is the standard form used in tight binding models of BN \cite{ribeiro2011a}. A similar electron-phonon Hamiltonian
without the static potential was considered by Covaci and Berciu
\cite{covaci2008a}.

Solution of this Hamiltonian is extremely involved, and complications
arise because electron-phonon interactions are retarded. The next
section describes how to solve the polaron problem for the graphene
lattice using DQMC.

\section{Method: Diagrammatic quantum Monte Carlo}
\label{sec:method}


I use the diagrammatic quantum Monte Carlo (DQMC) method to establish
the properties of polarons on the graphene lattice
\cite{prokofev1998a,mishchenko2000a}. In order to take account of the bipartite lattice and AB
modulated potential, the DQMC method has to be modified to include
basis. This is achieved by considering the
non-interacting Green function to have a matrix form. A slight
complication is presented by the off-diagonal terms of this matrix, which
are complex. This could in principle lead to a phase problem (which is
a generalized sign problem). The complex phase is found
to be very small when measuring the on-site Green functions of
interest here. For imaginary
time, $\tau_{f}>\tau_{i}$ and at absolute zero, the Green functions are defined
as follows:
\begin{eqnarray}
{\bf G} & = & \left( \begin{array}{cc}G_{AA} & G_{AB} \\ G_{BA} & G_{BB}\end{array} \right) \\
& = & \left( \begin{array}{cc}-\langle a(\tau_f)a^{\dagger}(\tau_i)\rangle & -\langle a(\tau_f)c^{\dagger}(\tau_i)\rangle \\ -\langle c(\tau_f)a^{\dagger}(\tau_i)\rangle & -\langle c(\tau_f)c^{\dagger}(\tau_i)\rangle\end{array} \right),
\end{eqnarray}
where,
%
\begin{eqnarray}
G_{AA} &=& \frac{\left[\exp(-E_{B} (\tau_{f} - \tau_{i}))+\exp(-E_{A} (\tau_{f} - \tau_{i}))\right]}{2}\\
& & + \frac{\Delta\left[\exp(-E_{A} (\tau_{f} - \tau_{i})) - \exp(-E_{B} (\tau_{f} - \tau_{i}))\right]}{ 2 \sqrt{|\phi_{\kvec}|^{2} + \Delta^{2}}}\nonumber
\end{eqnarray}

\begin{eqnarray}
G_{BB} &=& \frac{\left[\exp(-E_{B} (\tau_{f} - \tau_{i}))+\exp(-E_{A} (\tau_{f} - \tau_{i}))\right]}{2}\\
& & + \frac{\Delta\left[\exp(-E_{B} (\tau_{f} - \tau_{i})) - \exp(-E_{A} (\tau_{f} - \tau_{i}))\right]}{ 2 \sqrt{|\phi_{\kvec}|^{2} + \Delta^{2}}} \nonumber
\end{eqnarray}

\begin{equation}
G_{AB} = \frac{\phi_{\kvec}\left[\exp(-E_{B} (\tau_{f} - \tau_{i})) - \exp(-E_{A} (\tau_{f} - \tau_{i}))\right]}{ 2 \sqrt{|\phi_{\kvec}|^{2} + \Delta^{2}}}
\end{equation}

\begin{equation}
G_{BA} = \frac{\phi^{*}_{\kvec}\left[\exp(-E_{B} (\tau_{f} - \tau_{i})) - \exp(-E_{A} (\tau_{f} - \tau_{i}))\right]}{ 2 \sqrt{|\phi_{\kvec}|^{2} + \Delta^{2}}},
\end{equation}
%
$E_{A} = \mu + \sqrt{\Delta^2+|\phi_{\kvec}|^2}$ and $E_{B} = \mu -
\sqrt{\Delta^2+|\phi_{\kvec}|^2}$. Here, the pseudo chemical potential
$\mu$ allows greater control of the algorithm, but since a single
particle is simulated, the true chemical potential lies at the bottom
of the band. Thus, the results are only accurate when the electron
density is low (i.e. the system is doped well away from half-filling).
For $\tau_{f}<\tau_{i}$ and absolute zero, all Green functions are
zero valued because the polaron only contains a single electron. Since
$G(\tau_f<\tau_i)=0$, vertices are ordered between times $0$ and
$\tau$, where $\tau$ is the length of the diagram.

In its most basic form, the algorithm proceeds by inserting and
removing interaction lines into or from the electron propagator. The
propagators for the Holstein interaction with local phonons have the
form $\exp(-\omega_0\tau)\delta_{XX}$ where $X\in\{A,B\}$ represents
the sub-lattice type at the end of the propagator and $\delta_{XX}$ is
the Kronecker $\delta$-function. Sub-lattice type is fixed at the ends
of the whole diagram so that the dynamics of each symmetry broken
sub-lattice can be probed independently.

The imaginary time Green function tails off exponentially, and can
vary by several orders of magnitude, which makes direct measurement of
the Green function histogram impossible within a reasonable
time-scale. To avoid this, a Wang-Landau algorithm is used to make an
initial guess for the histogram, so that all diagram lengths, $\tau$
are visited a similar number of times during each simulation. The
advantage of the Wang-Landau algorithm is that it obtains the
histogram extremely fast. This histogram is not used directly for
computation of the Green function, because the bin size is finite
leading to systematic errors. Rather, it is used as input for a
reweighting procedure so that all imaginary times are visited
\cite{mishchenko2000a} and the $\tau$ dependent Green function is
calculated using the estimator given in
Ref. \onlinecite{mishchenko2000a} which corrects for finite histogram
bin size. Proper choice of the pseudo chemical potential, $\mu$,
speeds up the initialization of the Wang-Landau algorithm.

It is worth noting that $G_{AB}$ and $G_{BA}$ have a complex phase. A
Monte Carlo procedure can be obtained by keeping track of this phase
$e^{i\theta}$ such that averages of an estimator $O$ are given by $\langle e^{i\theta} O
\rangle / \langle e^{i\theta} \rangle$. For all cases considered here,
the average phase $\langle e^{i\theta} \rangle$ is found to be
extremely close to 1, and no expectation values had a complex
component after averaging. There is no obvious reason why the phase
should cancel (unlike in the 1D case where the signs exactly cancel
\cite{hague2011a}) and for large numbers of particles, the phase could
become a problem.

In order to obtain spectral functions, stochastic analytic inference
is used \cite{fuchs2010a}. Green functions are built up from
$\delta$-functions that can be moved continuously in frequency using a
separate Monte Carlo update scheme. The ability to construct spectral
functions from continuous frequencies is necessary to obtain reliable
analytic continuation at absolute zero where features can be very
sharply peaked. Each configuration of the spectral function is
weighted as $w\propto \exp(-\chi^2/2\alpha)$ and the factor $\alpha$ is
reduced from a large value until the average $\chi^2 < N_G$, the
number of points in the Green function. Averages are then
taken. Additional global updates (where all points can be shifted
simultaneously) have been included in the procedure to ensure that the
algorithm is ergodic.

\section{Results}
\label{sec:results}


This section begins by examining how the opening of a band-gap affects
the spectral functions when the electron-phonon coupling is switched
on. The phonon energy is set as $\hbar\Omega=t$ unless otherwise
specified. This very high value is chosen so that features relating to
polarons can be distinguished easily. Note that phonons of this energy
are still in the adiabatic regime at around one third of the half band
width. Naturally, this energy is much higher than that of any phonons
in graphene or BN, or of any surface phonons in the substrate, and
smaller values will also be discussed later in the article. Values for
$\Delta$ ranging from 0 to $t$ are also large for the same reason.
For simplicity (unless specified) the out-of-plane interaction
$g_{\kvec\qvec}$ is approximated to be momentum independent, and
$\tilde{g}=0$, leading to a local Holstein interaction.

Fig. \ref{fig:vardeltaimage} shows how the graphene spectral function, $A(E)$,
changes across the Brillouin zone. Spectral functions are computed from
the full Green functions, $\mathcal{G}_{AA}(\tau)$ and
$\mathcal{G}_{BB}(\tau)$, with $\mathcal{G}$ calculated on a
logarithmic mesh with 500 points. Separate image plots can be seen for
A and B type electrons to make the specific contributions from each
sub-lattice clear, and the non-interacting band structure is superimposed
for comparison. A moderate dimensionless electron-phonon coupling of
$\lambda = 0.7$ is chosen, with the exception of panel (b), which
shows spectral functions for the larger $\lambda=2.8$. Panels (a) and
(b) correspond to $\Delta=0$, (c) and (d) to $\Delta=t/2$ and (e) and
(f) to $\Delta=t$.

The main features of Fig. \ref{fig:vardeltaimage} are: (1) The
quasi-particle peaks are sharp at low energies $E-E_0 \lesssim
\hbar\Omega$ (where $E_0$ is the polaron ground state energy) but
broaden significantly for higher energies. This is especially
noticeable in panels (a), (d) and (f). (2) A clearly identifiable
polaron band (split off from the main dispersion) can be seen for
large $\lambda$ in panel (b). (3) Asymmetry between electrons on site
A and site B increases with $\Delta$. (4) The spectral gap can be seen
to increase with $\Delta$. The spectral gap is slightly smaller than
$\Delta$ due to broadening of the quasiparticle peak. (5) The
beginnings of a flat polaron band can be seen for $\Delta=t$ in panel
(f) and is just visible in panel (d) for $\Delta=t/2$. Again, it
should be noted that these results are for polarons (a single electron
at the bottom of an empty band interacting with phonon modes). As the
Fermi energy is changed, the spectral function close to the chemical
potential may be modified significantly. Far from the chemical
potential, differences should be less pronounced. Here, the spectral
gap is defined as the distance between the peaks of the spectral
function close to the Dirac point.

 When $\Delta=0$, A and B sites are symmetrical and the spectral
 function for each sub-lattice is identical, so results are only shown
 for the A sub-lattice in Figs. \ref{fig:vardeltaimage} (a) and
 (b). In the vicinity of the $\Gamma$ point, the spectral function is
 sharply peaked for states within an energy of around $\hbar\Omega$ of
 the bottom of the band, and quasi-particle lifetime (related to the
 inverse of the width) is greatly reduced for the highest energy
 states close to the tops of the bands. This is a polaronic effect:
 within an energy $\hbar\Omega$ of the bottom of the band the
 electrons have insufficient energy to excite real phonons. At the top
 of the band, the spectral functions are broad due to quasiparticle
 decays induced by interactions.

The spectral function is also sharply peaked close to the K point. By
examining the zero gap states (Fig. \ref{fig:vardeltaimage} (a) and
(b)) it is possible to consider whether it is possible that polaronic
states could simulate a gap. Examination of
Fig. \ref{fig:vardeltaimage}(a) shows that there is a flattening of
the dispersion near the K point (highlighted by comparison with the
non-interacting dispersion) accompanying a decrease in the width of
the quasi-particle peak. The flattening of the dispersion close to the
K point is associated with a steepening of the dispersion along the KM
and K$\Gamma$ lines, giving the dispersion the appearance of a
waterfall. The waterfall like features are at variance with the idea
that polaronic features could look like a band gap
\cite{rotenberg2008a}, since emulation of a gap would require the
dispersion to show the opposite shape to that seen here: a very steep
band structure close to K with a rapid change of gradient leading to
flattening between K and M.  Panel (b) shows that the spectral weight
gets broader with increased $\lambda$, indicating that it is also
likely that finite quasi-particle lifetimes contribute to an obscuring
of any gap rather than any apparent gap opening. Since the results are
taken with the chemical potential at the bottom of the band, the
results are speculative and further studies at higher densities would
be required to reach a full conclusion.


A major feature in panel (b) for $\lambda=2.8$ is the emergence of a
flat polaron band, which separates off from the lower band. Such a
feature typically corresponds to the energy required to excite a real
phonon, and is singular in basic Migdal--Eliashberg theories of
electron-phonon interactions (see
e.g. Ref. \onlinecite{nakajima1980a}). The large electron-phonon
coupling leads to a significant decrease in the quasi-particle
lifetime at large energies, however, the band flattening seen around
the K point (which is unconnected to the polaron band) persists. The
ground state polaron energy is lowered due to the polaron
self-interaction (seen as the offset from the non-interacting band at
the $\Gamma$ point).

Increase in $\Delta$ breaks the symmetry between A and B sub-lattices,
and this can be seen in Fig. \ref{fig:vardeltaimage} panels (c)-(f). A
gap opens as $\Delta$ is increased. A band at high energies, roughly
tracing the dispersion of the ungapped, non-interacting band, can be
seen in panels (d) and (f), although its spectral weight is extremely
small. The origins of this band are unclear. Increased $\Delta$ also
leads to a flat polaron band, that separates from the main band (it
can be seen as the lowest energy feature at the K point). This feature
is only just visible in panel (d), but is clearly separated from the
main band in panel (f). The appearance of this band at weak $\lambda$
is a consequence of polaron localization at large $\Delta$, which
increases polaron self-interaction (i.e. the effective $\lambda$ is
increased by the localization).

To demonstrate that the effects are not artifacts of the local
Holstein interaction, Fig. \ref{fig:vardeltaimagefroh} shows the
spectral functions computed for interactions that include nearest
neighbor forces, the leading correction to the Fr\"ohlich interaction
(N.B. Sums associated with the Fourier transforms of the force
functions are truncated so that they are not prohibitively
computationally expensive. A truncated near neighbor interaction
contains sufficient physics to obtain good agreement with the full
interaction \cite{hague2009a,hague2010a}). The plots are qualitatively
similar to those in Fig. \ref{fig:vardeltaimage}. The main difference
is that the longer range forces slightly reduce the effects of
interactions at a particular $\lambda$: for the Fr\"ohloich
interaction, there is slightly less drop in the ground state energy of
the polaron band, and the flat features appear at slightly higher
energies relative to the bottom of the band. The slight increase in
the width of the polaron band occurs because the distortions
associated with the Fr\"ohlich polaron are already preformed before
the electron hops between sites, so the effective mass is smaller. The
differences between Holstein and Fr\"ohlich interactions are more
pronounced for larger interaction strengths, as can be seen by
comparing the $\lambda=2.8$ results.

To clarify the effects of interaction and bare gap on the spectral
function, Fig. \ref{fig:vardelta} shows the same data, but with both
spectral function types superimposed on the same plot. For clarity,
spectral weight below a cutoff of less than 0.1 is not shown
(variations on this order of magnitude are not distinguishable below
this scale, and the resulting curve appears as a series of straight
horizontal lines obscuring the plots). From top to bottom $\Delta=0$,
$\Delta=t/2$ and $\Delta=t$ and $\lambda=0.7$ in all panels. The gap
opening can clearly be seen. The effects of interaction, which
increases the width of the spectral function at large energies (where
real phonons can be created) is much clearer in these plots.

Figs. \ref{fig:varlambdaimage} and \ref{fig:varlambda} show how the
graphene spectral function changes across the Brillouin zone when the
electron-phonon coupling is varied and $\Delta=t$ (from top to bottom,
$\lambda=0.7, 1.4, 2.1$ and $2.8$). Spectral functions for the A
sub-lattice can be seen on the left of Fig. \ref{fig:varlambdaimage}
(panels (a), (c), (e), (g)) and spectral functions for the B
sub-lattice are shown on the right hand side of the plot (panels (b),
(d), (f) and (h)). Self-interactions lower the polaron energy, and it
can be seen that spectral weight at the $\Gamma$ point moves to lower
energies relative to the non-interacting bands as $\lambda$
increases. At large $\lambda$, a sharply defined polaron band
separates from the main band, and can be seen as a low energy band
that is almost flat and has a high quasi-particle lifetime (again,
this is the singular band found in conventional Migdal--Eliashberg
theories). The remnant of the non-interacting band is visible as a
side-band above the polaron band. The quasi-particle lifetime
decreases dramatically with increased $\lambda$, seen as a broadening
of the spectral function. A shadow band can be seen in the spectral
functions for B sites, although it is relatively weak, and decreases
in weight as $\lambda$ is increased. Examination of
Fig. \ref{fig:varlambda} (which shows spectral functions for both
sub-lattices superimposed onto the same plot), shows that the spectral
gap is robust against increase in $\lambda$, although a number of
additional excitations appear. Again, for completeness, the effects of
longer range interactions are shown in
Fig. \ref{fig:varlambdaimagefroh}. For the larger bare band gaps, the
effects of interaction range are extremely small for all values of
$\lambda$.

Finally, Fig. \ref{fig:diracvarlambda} shows how the spectral
function at the K point evolves as the electron-phonon coupling is
increased. Three main peaks are visible: (1) The lowest corresponds to
the polaron band, (2) electrons on the A sub-lattice have at least 1
excited state, and (3) B type electrons can be seen at the highest
energies.

In the strongly doped system, the gap between the A excited state and
the B electron energy is essentially unchanged by an increase in the
electron-phonon coupling, with the gap remaining robust. The polaron
band can be seen splitting off from the non-interacting band, with the
polaron energy dropping rapidly at small $\lambda$, followed by a
sustained decrease. This is related to the flattening of the polaron
band seen in Fig. \ref{fig:varlambdaimage}. As the electron-phonon
coupling is increased, the quasi-particle lifetime drops, seen as a
broadening of the peak. There is an increase in the energy difference
between the bottom of the A band and the bottom of the B band at the
Dirac point, indicating an increase in the potential barrier formed by
the higher energy A sites. This indicates an increased transport gap
at the K point from electron-phonon interactions, consistent with
perturbation theory calculations at half-filling
\cite{hague2011a}. Here, transport gap is taken to be the difference
between the lowest energy states on A and B sub-lattices.

Finally, I consider the effects of lower phonon frequencies,
dispersive phonons and smaller bare band gaps in
Fig. \ref{fig:dispersivephonons}. In the figure, $\lambda=2.8$
throughout. The truncated Fr\"ohlich interaction is used in all
panels. Comparison of panels (a) and (b) with panels (c) and (d) shows
that including dispersive phonons with a half band width of approximately
30\% of the phonon frequency has no qualitative effect on the spectral
functions. Far more important is the decrease in phonon frequency. At
this large value of lambda, the sharp flat feature dissipates, and all
features are generally broadened (panels (e) and (f)). Again, the
inclusion of phonon dispersion causes no major changes in the spectral
function.

\section{Summary and conclusions}
\label{sec:conclusions}

In this paper, I have used the diagrammatic quantum Monte Carlo
technique to compute the spectral functions of polarons on a honeycomb
lattice in the presence of a substrate or superstrate. Extensions to
DQMC were introduced to deal with the bi-partite honeycomb (graphene)
lattice. Spectral functions were obtained for a variety of
electron-phonon coupling strengths and substrate induced sub-lattice
symmetry breaking using stochastic analytic inference. The results
presented here relate to heavily doped graphene.

Electron-phonon interactions are seen to have a range of effects on
the band structure: The quasi-particle peaks are sharp at low energies
$E-E_0 < \hbar\Omega$, but broaden significantly for higher energies. A
clearly identifiable polaron band with large quasi-particle lifetime
forms. Asymmetry between electrons on different sub-lattices increases
with $\Delta$.

Flattening of the band around the K point in the absence of
sub-lattice symmetry breaking suggests that electron-phonon coupling
could not emulate a gap in the absence of a modulated
potential. Spectral gaps induced by substrates are seen to be
reasonably robust against interactions at heavy doping, although a
shortening of quasi-particle lifetime (broadening of the spectral
function) may make the gap difficult to discern.

A gap in the spectral function is induced on increase of the energy
difference between sub-lattices, $\Delta$, but is slightly reduced by
broadening of the quasiparticle peak at large $\lambda$. The formation
of a polaron band on increase of electron-phonon coupling increases
the transport gap at the K point due to an increase in the energy
difference between sub-lattices. This provides additional evidence
that strongly polarizable substrates and superstrates could be used to
enhance transport gaps opened by a substrate. Further work is underway
to examine spectral functions and gaps close to half filling.

\section{Acknowledgments}

I am pleased to acknowledge EPSRC grant EP/H015655/1 for funding and useful discussions with P.E. Kornilovitch, A.S. Alexandrov, M. Roy, P. Maksym, E. McCann, V. Fal'ko, E. Burovski, N.J. Mason, N.S. Braithwaite and A. Davenport.

\bibliography{graphene_arpes}

\end{document}